\documentclass{article}
\pdfoutput=1
\usepackage{graphicx}
\usepackage{amsmath}
\usepackage{url}
\usepackage[colorlinks=true,linkcolor=blue,anchorcolor=black,citecolor=blue,filecolor=black,menucolor=black,urlcolor=blue,breaklinks=true,pdfhighlight=/P,pdfmenubar=true,pdftoolbar=true,pdfpagelabels=true,pdfstartpage=1,pdfstartview=FitV,pdftitle={A Round-Robin Tournament of the Iterated Prisoner's Dilemma with Complete Memory-Size-Three Strategies},pdfsubject={A Round-Robin Tournament of the Iterated Prisoner's Dilemma with Complete Memory-Size-Three Strategies},pdfauthor={Tobias Kretz},pdfcreator={Tobias Kretz},pdfproducer={Tobias Kretz},pdfkeywords={prisoners dilemma}]{hyperref}
\usepackage[numbers,sort&compress]{natbib}
\usepackage{hypernat}
\usepackage[english]{babel}

\begin{document}

\title{A Round-Robin Tournament of the Iterated Prisoner's Dilemma with Complete Memory-Size-Three Strategies}
\author{Tobias Kretz\\
{PTV -- Planung Transport Verkehr AG}\\
{Stumpfstra{\ss}e~1}\\
{D-76131 Karlsruhe}\\ 
{Germany}\\
}

\maketitle

\begin{abstract}
In this paper the results of a simulation of a prisoner's dilemma robin-round tournament are presented. In the tournament each participating strategy plays an iterated prisoner's dilemma against each other strategy (round-robin) and as a variant also against itself. The participants of a tournament are all strategies that are deterministic and have the same size of memory with regard to their own and their opponent's past actions: up to three most recent actions of their opponent and up to two most recent actions of their own. A focus is set on the investigation of the influence of the number of iterations, details of the payoff matrix, and the influence of memory size. The main result is that for the tournament as carried out here, different strategies emerge as winners for different payoff matrices, even for different payoff matrices being similar judged on if they fulfill relations $T + S = P + R$ or $2R > T + S$. As a consequence of this result it is suggested that whenever the iterated prisoner's dilemma is used to model a real system that does not explicitly fix the payoff matrix, one should check if conclusions remain valid, when a different payoff matrix is used.
\end{abstract}

\section{Introduction and Motivation}
The prisoner's dilemma \cite{Flood1958,_Axelrod1985} is probably the most prominent and most discussed example from game theory, which is a result of it standing as {\em the} model of the formation of co{\"o}peration in the course of biological as well as cultural evolution \cite{_Axelrod1985,Axelrod1981}. 

A na{\"i}ve interpretation of Darwin's theory might suggest evolution favoring nothing but direct battle and plain competition. However numerous observations of co{\"o}peration in the animal kingdom oppose this idea by plain evidence. While such examples among animals are impressive in itself, clearly the most complex and complicated interplay of co{\"o}peration and competition occurs with humans; a fact which becomes most obvious when a large number of humans gathers as a crowd in spatial proximity. There are astonishing and well-known examples for both: altruism among strangers under dangerous external conditions \cite{Sime1980,Keating1982,Laur1997,Quarantelli2001,Clarke2002,Mawson2005,Fahy2005,Drury2009} as well as fierce competition on goods with very limited material value often linked with a lack of information \cite{McFadden1991,Schelajew2000} -- and anything in between these two extremes; see for example the overviews in \cite{Kretz2007a,Schreckenberg2008}. In relation to these events -- and possible similar events of pedestrian and evacuation dynamics \cite{Schadschneider2009} to come in the future -- the wide-spread na{\"i}ve interpretation of the theory of evolution in a sense poses a danger, as it might give people in such situations the wrong idea of what their fellows surrounding them are going to do and by this in turn suggest overly competitive and dangerous behavior. Knowing of said historic events together with having an idea of theories that suggest why co{\"o}peration against immediate maximal self-benefit can be rational hopefully can immunize against such destructive thoughts and actions.

From the beginning the prisoner's dilemma was investigated in an iterated way \cite{_Rapoport1965,Trivers1971}, often including that the ability of strategies to hark back on course of events of the tournament \cite{_Axelrod1985,_Axelrod1997} was unlimited, i.e. they had a memory potentially including every own and opponents' steps. Despite the possibility of using more memory the first strategy emerging as winner -- {\em tit-for-tat} -- did with a memory of only the most recent action of the opponent. Another famous and successful strategy -- {\em pavlov} -- also makes only use of a small memory: it just needs to remember its own and the opponent's action. In this contribution the effect of an extension of the memory up to the three latest actions of the opponent and up to two latest own actions is investigated.

In the course of discussion of the prisoner's dilemma a number of methods have been introduced like probabilistic strategies to model errors (``noise'') \cite{Nowak1993}, evolutionary (ecologic) investigation \cite{_Axelrod1985}, spatial relations (players only play against spatially neighbored opponents) \cite{Baek1989,Nowak1992,Nowak1993b,Grim1997,Nakamaru1997,Kirchkamp2000,Schweitzer2002,Masuda2003,Fort2005,Alonso2006}, and creation of strategies by genetic programming \cite{Axelrod1981,Nowak1993,Holland1992,Michalewicz1996,Salhi1996}. Most of these can be combined. For an overview on further variants see review works like \cite{Doebeli2005,Kuhn2007}.

Contrary to these elaborate methods, a main guiding line in this work is to avoid arbitrary and probabilistic decisions like choosing a subset of strategies of a class or locating strategies spatially in neighborhoods; spatial variants as well as a genetic approach are excluded. Instead each strategy of the class participates and plays against each other. A consequence from investigating complete classes is that it is impossible to have a continuous element as constructing element of a strategy; this forbids probabilistic strategies. The round-robin mode as well -- at least in parts -- is a consequence of avoiding arbitrariness: drawing lots to choose pairs of competitors like in tournaments would bring in a probabilistic element. In other words: the source code written for this investigation does not at any point make use of random numbers. It is a deterministic brute force calculation of a large number of strategies and a very large number of single games. The relevance lies not in modeling a specific system of reality, but in the completeness of the investigated class and in general the small degree of freedom (arbitrariness) of the system. 

By the strictness and generality of the procedure, a strategy can be seen as a Mealy automaton or the iterative game between two strategies as a Moore machine \cite{Moore1956,Abreu1988,Linster1992,Miller1996} respectively a spatially zero-dimensional cellular automaton \cite{_Wolfram1986,_Wolfram2002} (see section \ref{sec:ca}).

\section{Definition of a Strategy}
In the sense of this paper a strategy with a memory size $n$ has $n+1$ sub-strategies to define the action in the first, second, ... $n$th and any further iteration. The sub-strategy for the first iteration only decides, how a strategy starts the tournament, the sub-strategy for the second iteration depends on the action(s) of the first iteration, the sub-strategy for the third iteration depends on the actions in the first and second iteration (if memory size is larger one) and the sub-strategy for the $(N>n)$th iteration depends on the actions in the $(N-n)$ to $(N-1)$st iteration (compare Figure \ref{fig:example}).

A similar approach has been followed in \cite{Beaufils1998}, but there are differences in the definition of the class concerning the behavior in the first $n-1$ iterations and most important it has not been used for a round-robin tournament with all strategies of a class, but combined with a genetic approach.

Another investigation dealing with effects of memory size is \cite{Hauert1997}. The difference there is that the strategies are probabilistic and (therefore) not all strategies participate in the process.

\subsection{Data Size of a Strategy, Number of Strategies, and Number of Games}
\begin{figure}[htb]
  \begin{center}
    \includegraphics[width=0.55\textwidth]{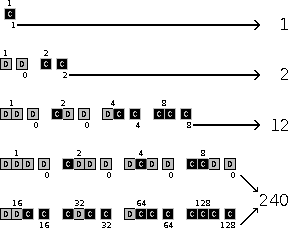}
  \end{center}
    \caption{Tit-for-tat as strategy (1/2/12/240). The part (1/2/12) applies only in the starting phase, when only no, one or two earlier states of the opponent exist. So, co{\"o}peration is coded with a ``1'', defection with a ``0''. If a strategy remembers also its own past actions then these are always stored in the lower bits, i.e. for example of the triples, the leftmost would indicate a strategy's own preceding action and the middle and right would indicate the second to last and last action of the opponent (``low to high'' is ``left to right'').
} \label{fig:example}
\end{figure}

Since at the beginning there is no information from the opponent a strategy consists of a decision, how to begin an iterated game (one bit). In the second round, there is only information on one past step from the opponent, so the strategy includes the decision how to react on this (two bits), the third step is still part of the starting phase and therefore also has its own part of the strategy (four bits, if the decision does not depend on a strategy's own preceding action). Therefore there are 128 strategies if there is a no-own-two-opponent memory. Finally with size-three memory, one has eight more bits. As an example in Figure \ref{fig:example} it is shown, how one calculates the number combination (1/2/12/240) from the tit-for-tat strategy. These 15 bits lead to a total of $N=32768$ different strategies. If each strategy plays against each other strategy and against itself one has to calculate $N\cdot(N+1)/2=2^{29}$ different iterated prisoner's dilemmas. 

Table \ref{tab:memorysizes} sums up these numbers for different memorysizes. To remember the last $n$ actions of a pair of strategies, one needs $2n$ bits and for the results of a strategy over the course of iterations one needs -- depending on the kind of evaluation -- a few bytes for each pair of strategies. The number of pairs of strategies -- and this is the limiting component -- grows at least approximately like $2^{2^{n+2}-3}$. On today's common PCs RAM demands are therefore trivial up to a memory size of $n=2$, in the lower range of 64-bit technology (some GBs of RAM) for $n=3$, and totally unavailable for $n=4$ and larger (more than an exabyte).

\begin{table}
\begin{center}
\begin{tabular}[htbp]{crrr}
Memory size & \#Bits & \#Strategies  & \#Games in one iteration \\
self /other &        &               &         \\ \hline 
0 / 0           & 1  &             2 & 1 resp. 3 \\
0 / 1           & 3  &             8 & 28 resp. 36 \\            
1 / 1           & 5  &            32 & 496 resp. 528 \\            
0 / 2           & 7  &           128 & 8,128 resp. 8,256 \\
1 / 2           & 13 &         8,192 & $\approx 33.55 \cdot 10^{6}$ \\            
2 / 1           & 13 &         8,192 & $\approx 33.55 \cdot 10^{6}$ \\            
0 / 3           & 15 &        32,768 & $\approx 536.8 \cdot 10^{6}$ \\
2 / 2           & 21 &     2,097,152 & $\approx 2.199 \cdot 10^{9} $ \\            
1 / 3           & 29 &   536,870,912 & $\approx 144.1 \cdot 10^{15}$ \\            
3 / 1           & 29 &   536,870,912 & $\approx 144.1 \cdot 10^{15}$ \\            
0 / 4           & 31 & 2,147,483,648 & $\approx 2.306 \cdot 10^{18}$\\
\end{tabular}
\end{center}
\caption{Number of bits ($b$) to represent a strategy, number of strategies ($2^b$), and number of prisoner's dilemma games in an iteration step in a round-robin tournament ($2^{b-1}(2^b\pm 1)$) for different memory sizes. This leads to a computational effort shown in Table \ref{tab:computing}. }
\label{tab:memorysizes}
\end{table}

\begin{table}
\begin{center}
\begin{tabular}[htbp]{crc}
Memory size & RAM    & Time          \\
self /other &        &               \\ \hline 
0 / 0       &   10 B & insignificant \\
0 / 1       &  100 B & insignificant \\            
1 / 1       &  10 KB & s .. min      \\            
0 / 2       & 100 KB & s .. min      \\
1 / 2       & 100 MB & min .. d      \\            
2 / 1       & 100 MB & min .. d      \\            
0 / 3       &  10 GB & h .. weeks    \\
2 / 2       &  10 TB & d .. year     \\            
1 / 3       &   1 EB & $>$ year      \\            
3 / 1       &   1 EB & $>$ year      \\            
0 / 4       &  10 EB & decade(s) (?) \\
\end{tabular}
\end{center}
\caption{Magnitudes of computational resource requirements (on a double quad core Intel Xeon 5320). The computation time depends significantly on the number of different payoff matrices to be investigated. Large scale simulations with parallel computing of the iterated prisoner's dilemma has also been dealt with in \cite{Townsley2006}.}
\label{tab:computing}
\end{table}

\section{The Cellular Automata Perspective} \label{sec:ca}
This section serves to have another perspective at the system in terms of cellular automata. This can help to get a visual idea of the system dynamics. However, the reader may well skip this and proceed to the next section.

\begin{figure}[htbp]
  \begin{center}
    \includegraphics[width=0.55\textwidth]{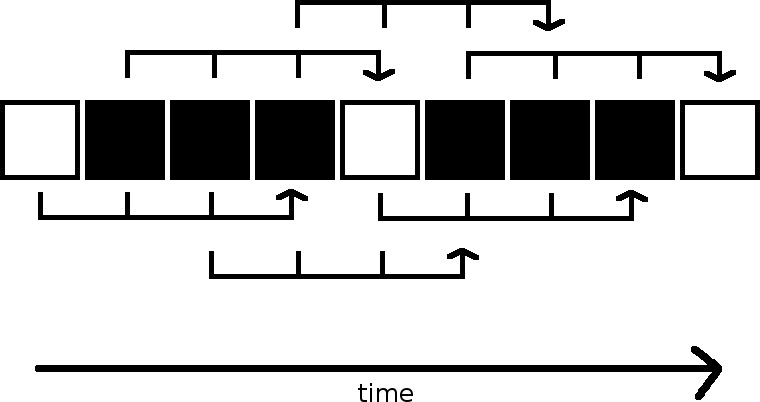}
  \end{center}
    \caption{Rule 110 applied self-referentially to a point-like cellular automaton with memory. Note: as time increases toward the right and the most recent state is meant to be stored in the highest bit, but higher bits are notated to the left, one has to reverse the bits compared to Wolfram's standard notation.} 
    \label{fig:ca-demo1}
\end{figure}

\begin{figure}[htbp]
  \begin{center}
    \includegraphics[width=0.55\textwidth]{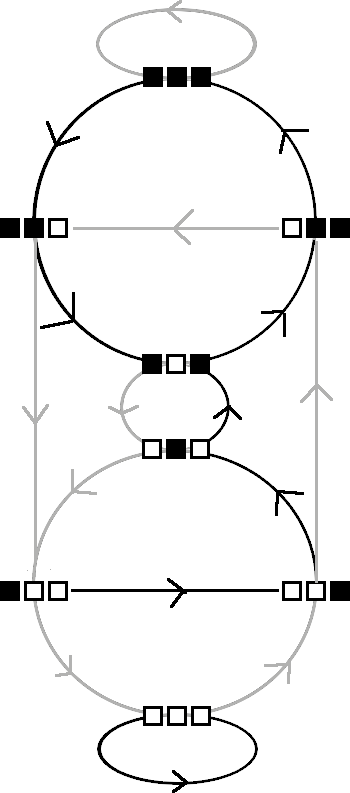}
  \end{center}
    \caption{Transition graph for rule 110 (black links) and possible links or other rules (grey links).} 
    \label{fig:transitions}
\end{figure}

\begin{figure}[htbp]
  \begin{center}
    \includegraphics[width=0.55\textwidth]{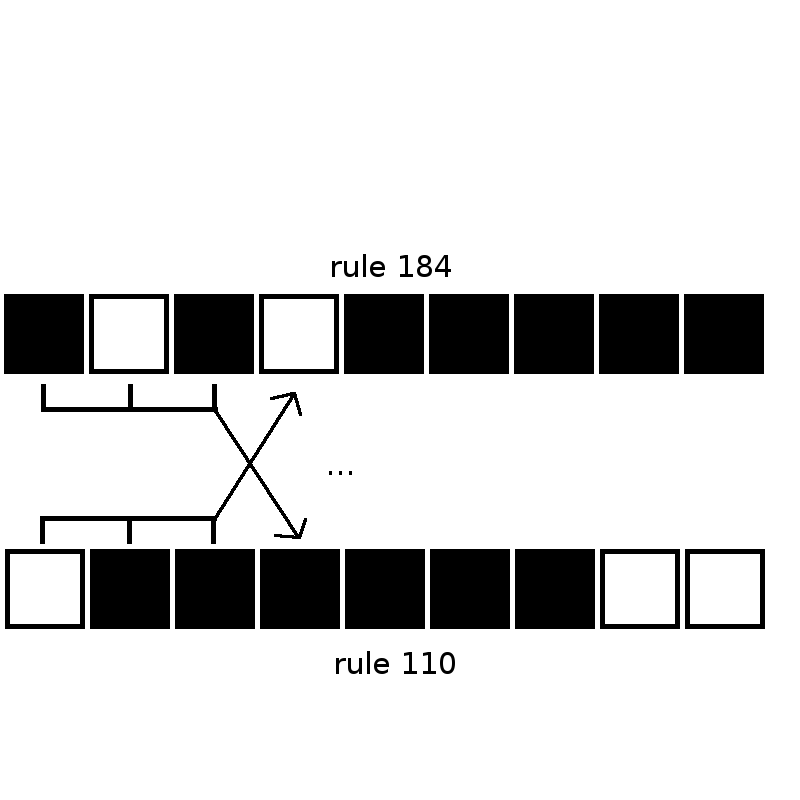}
  \end{center}
    \caption{Rule 184 and rule 110 interacting. As a model for the iterated prisoner's dilemma the dependence here models the situation that a prisoner remembers the three preceding moves of the opponent but none of its own.} 
    \label{fig:ca-demo2}
\end{figure}

\begin{figure}[htbp]
  \begin{center}
    \includegraphics[width=0.55\textwidth]{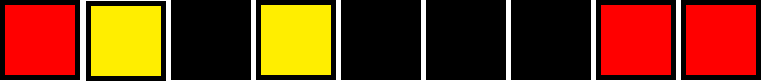}
  \end{center}
    \caption{Figure \ref{fig:ca-demo2} as one single cellular automaton. If the states of both automata are white (black) the state here is shown as well as white (black). If 184 is white (black) and 110 black (white), the state here is yellow (red).} 
    \label{fig:ca-demo3}
\end{figure}

Wolfram's elementary cellular automata are defined (or interpreted) to exist in one spatial plus one temporal dimension. However, one can also apply the rules to a point-like cellular automaton with memory.  Figure \ref{fig:ca-demo1} shows an example for this.
One can also interpret this system not as a cellular automaton that has a memory and a binary state, but as an automaton that can have one of eight states with transitions between the states being restricted. For the full set of 256 rules each state can be reached in principle from two other states and from a particular state also two states can be reached. Choosing a specific rule is selecting one incoming and one outgoing state. This is exemplified in Figure \ref{fig:transitions} for rule 110.
For the iterated prisoner's dilemma one needs two such cellular automata that interact that determine their next state from the data of the other automaton as shown in Figure \ref{fig:ca-demo2}.
It is of course possible to interpret two interacting cellular automata as one single point-like cellular automaton with a larger set of states. Then Figure \ref{fig:ca-demo2} would translate to Figure  \ref{fig:ca-demo3}.
One now again could draw a transition graph (with 64 nodes that all have one of four possible incoming and outgoing links or a specific combination of rules) for further theoretical analysis. For this work we shall now abandon these basic and theoretical considerations and just adhere to the fact that the implementation of the process can be seen as a cellular automaton, more precisely an enormous number of combination of interacting very simple cellular automata.

\section{Payoff Matrix}
\begin{table}
\begin{center}
\begin{tabular}[htbp]{c||c|c}
     & C(2)      &  D(2)  \\ \hline \hline
C(1) & $R$ \ $R$ & $S$ \ $T$ \\ \hline
D(1) & $T$ \ $S$ & $P$ \ $P$ \\
\end{tabular}
 \end{center}
\caption{General payoff matrix}
\label{tab:payoffmatrix}
\end{table}

The four values $T$, $R$, $P$, and $S$ of the payoff matrix (see Table \ref{tab:payoffmatrix}) need to fulfill the relation 
\begin{equation}
T > R > P > S \label{eq:basic}
\end{equation}
to be faced with a prisoner's dilemma. For the purpose of this contribution one can choose $S=0$ without loss of generality, as whenever the payoff matrix is applied all strategies have played the same number of games. In addition to equation \ref{eq:basic} it is often postulated that
\begin{equation}
2R > T \label{eq:2RT}
\end{equation}
holds.

The equation
\begin{equation}
T + S = P + R\label{eq:TSPR}
\end{equation}
as well marks a special set of payoff matrices, as those values can be seen as model of a trading process, where the exchanged good has a higher value for the buyer $i$ than the seller $j$:
\begin{equation}
p_{ij}=\alpha+\beta\delta_i-\gamma\delta_j \label{eq:Price1}
\end{equation}
where $\delta=1$, if a player co{\"o}perates and $\delta=0$ if he defects. Therefore $\beta$ can directly be interpreted as the ``gain from receiving'' value and $\gamma$ the ``cost from giving'' value. $\alpha$ is a constant to guarantee $p_{ij}\geq 0$. $T$, $R$, $P$ for technical convenience, and $S$ can be calculated from these: $T=\alpha+\beta$, $R=\alpha+\beta-\gamma$, $P=\alpha$, and $S=\alpha-\gamma$. Aside from the descriptive interpretation as ``gain from receive'' and ``cost to give'' this reparametrization has the advantage that the original condition equation (\ref{eq:basic}) and the additional conditions equation (\ref{eq:2RT}) and $S=0$ reduce to $\beta > \gamma=\alpha$. Furthermore it is the form of the basic equation in George Price's model for the evolution of co{\"o}peration \cite{Price1970,Frank1995}.

As we do not only want to investigate payoff matrices, where equations (\ref{eq:2RT}) and (\ref{eq:TSPR}) hold, we rewrite
\begin{eqnarray}
T &=& (1+a+b)P\\
R &=& (1+a)P\\
a&=&\alpha/P-1>0\\
b&=&\beta/P>0
\end{eqnarray}
In principle one could set $P=1$ without loss of generality, but then it was not possible to write all combinations holds/does not hold of equations (\ref{eq:2RT}) and (\ref{eq:TSPR}) with integer-valued $T$ and $R$.
Now equation (\ref{eq:TSPR}) simply can be written as
\begin{eqnarray}
b=1
\end{eqnarray}
and shall be investigated as one variant next to $b>1$ and $b<1$. And equation (\ref{eq:2RT}) writes
\begin{eqnarray}
a + 1 > b.
\end{eqnarray}
Here as well $a+1=b$ and $a+1<b$ will be investigated (always taking care that $a>0$ and $b>0$ hold). Finally, $a (<,=,>) 1$ and $a (<,=,>) b$ are relevant conditions, if it's possible to distinguish in this way.

Obviously not all combinations of these conditions can hold simultaneously. ($a+1<b$, $b<1$) for example has no allowed solution. The allowed combinations and the values for $T$, $R$, and $P$ are shown in Table \ref{tab:TRP}. For each combination of conditions an infinite number of values could have been found. One could have chosen to interpret ``$>$'' as ``much greater than'' but then selecting specific numbers in a way would have been arbitrary. So the smallest numbers to fulfill a set of conditions have been chosen as representatives.

\begin{table}
\begin{center}
\begin{tabular}[htbp]{ccc|ccc|cc}
Cond. 1     & Cond. 2     &  Cond. 3     & $T$ & $R$ & $P$ & $T = R+P$ & $2R>T$\\ \hline
$b = 1     $& $ a = 1    $&              &  3  &  2  &  1  & holds & holds\\
$b = 1     $& $a > 1     $&              &  4  &  3  &  1  & holds & holds\\
$b = 1     $& $    a < 1 $&              &  5  &  3  &  2  & holds & holds\\
$    b < 1 $& $a = 1     $&              &  5  &  4  &  2  &       & holds\\
$    b < 1 $& $a > 1     $&              &  6  &  5  &  2  &       & holds\\
$    b < 1 $& $    a < 1 $&$ b = a      $&  4  &  3  &  2  &       & holds\\
$    b < 1 $& $    a < 1 $&$ b > a      $&  6  &  4  &  3  &       & holds\\
$    b < 1 $& $    a < 1 $&$ b < a      $&  6  &  5  &  3  &       & holds\\
$b > 1     $& $b < a + 1 $&$ a > 1      $&  5  &  3  &  1  &       & holds\\
$b > 1     $& $b < a + 1 $&$ a = 1      $&  7  &  4  &  2  &       & holds\\
$b > 1     $& $b < a + 1 $&$     a < 1  $&  9  &  5  &  3  &       & holds\\
$b > 1     $& $b = a + 1 $&$ a = 1      $&  4  &  2  &  1  &       &    \\
$b > 1     $& $b > a + 1 $&$ a = 1      $&  5  &  2  &  1  &       &    \\
$b > 1     $& $b = a + 1 $&$ a > 1      $&  6  &  3  &  1  &       &    \\
$b > 1     $& $b > a + 1 $&$ a > 1      $&  7  &  3  &  1  &       &    \\
$b > 1     $& $b = a + 1 $&$     a < 1  $&  6  &  3  &  2  &       &    \\
$b > 1     $& $b > a + 1 $&$     a < 1  $&  7  &  3  &  2  &       &    \\
\end{tabular}
  \end{center}
\caption{Investigated variants of values for the payoff matrix.}
\label{tab:TRP}
\end{table}

\section{Iteration, Tournament, and Scoring}
\label{sec:score}
In an iteration step all strategies play a prisoner's dilemma against any of the other strategies and themselves. For this a strategy calculates its action from the preceding actions of the specific opponent. If $N_{ij}^t$, $N_{ij}^r$, $N_{ij}^p$, $N_{ij}^s$ are the counters, how often strategy $i$ received a $T$, $R$, $P$ or $S$ payoff playing against a specific strategy $j$, in each iteration step for each $i$ and each $j$ one of the four $N_{ij}^{x}$ is increased by one.

Now all the payoff matrices from Table \ref{tab:TRP} are applied one after the other to calculate for each payoff matrix for each strategy $i$ the total payoff $G_i^1$:
\begin{equation}
G_i^1= \sum_j T N_{ij}^T + R N_{ij}^R + P N_{ij}^P
\end{equation}

The strategy (or set of strategies) $i$ yielding the highest $G_i^1$ is one of the main results for a specific iteration round and a specific payoff matrix.

Then the tournament is started. Each tournament round $g$ is started by calculating the average payoff of the preceding tournament round:
\begin{equation}
G^{g}=\frac{\sum_i G_i^{g}\delta_i^{g}}{\sum_i \delta_i^{g}}
\end{equation}
where $\delta_i^{g}=1$, if strategy $i$ was still participating in the tournament in tournament round $g$ and $\delta_i^{g}=0$ else. Then $\delta_i^{g+1}$ is set to 0, if $\delta_i^{g}=0$, or if a strategy scored below average:
\begin{equation}
G_i^{g}<G^{g}
\end{equation}
The payoff for the next tournament round $g+1$ is calculated then for all strategies still participating in the tournament:
\begin{equation}
G_i^{g+1}= \sum_j (T N_{ij}^T + R N_{ij}^R + P N_{ij}^P)\delta_j^{g+1}
\end{equation}
The tournament ends, if only one strategy remains or if all remaining strategies score equal in a tournament round (i.e. they have identical $G_i^{g}$). The strategies, which manage to emerge as winners of such a tournament are the second main result for a specific iteration step and a specific payoff matrix.

Such an elimination tournament can be interpreted as an evolutionary tournament, where the frequency values for the strategies can only take the values $f=0$ and $f=1$.

To state it explicitly: all strategies participate again in the next iteration step for another first round of the tournament. The elimination process only takes place within an iteration step and not across iteration steps, and there is no prisoner's dilemma game played in or between the rounds of a tournament. As all strategies are deterministic this procedure is equivalent to playing the prisoner's dilemma a fixed number of iterations, evaluate the scores, eliminate all strategies scoring below average and play again the fixed number of iterations with the remaining strategies, and so on.

\section{Results}
In this section for all payoff matrices of Table \ref{tab:TRP} the strategies are given that for large numbers of iteration steps have the highest payoff $G_i^1$ in the first round of the tournament and those strategies that win the tournament -- if the system stabilizes to one winner. Additionally the iteration round is given, when this winning strategy (strategies) appeared for the first time to stay continuously until the last calculated iteration. This implies that for a certain payoff matrix prior to this iteration the number of iterations is important for the question which strategy will emerge as the best (in the sense described in section \ref{sec:score}).

\subsection{Results for No-Own-One-Opponent Memory}
With only one action to remember, there are just 8 strategies (named (0/0) to (1/3)). (0/0) never co{\"o}perates, (1/3) always. TFT is (1/2). 1000 iteration steps were done. It's safe to say that this is sufficiently long, as the results -- shown in tables \ref{tab:1-0} and \ref{tab:1-1} -- stabilize at latest in iteration step 16 (respectively 179). 

\begin{table}
\begin{center}
\begin{tabular}[htbp]{ccc|c|c|c}
$T$ &$R$&$P$& First it. &  $G_i^1$ &  Tournament \\ \hline
3 & 2 & 1 & 8 & (0/0) & (1/2)\\ 
4 & 3 & 1 & 4 & (0/0) & (1/2)\\ 
5 & 3 & 2 & 16 & (0/0) & (1/2)\\ 
5 & 4 & 2 & 6 & (0/0) & (1/2)\\ 
6 & 5 & 2 & 4 & (0/0) & (1/2)\\ 
4 & 3 & 2 & 10 & (0/0) & (1/2)\\ 
6 & 4 & 3 & 14 & (0/0) & (1/2)\\ 
6 & 5 & 3 & 6 & (0/0) & (1/2)\\ 
5 & 3 & 1 & 4 & (0/0) & (0/0)\\ 
7 & 4 & 2 & 4 & (0/0) & (0/0)\\ 
9 & 5 & 3 & 4 & (0/0) & (0/0)\\ 
4 & 2 & 1 & 4 & (0/0) & (0/0)\\ 
5 & 2 & 1 & 4 & (0/0) & (0/0)\\ 
6 & 3 & 1 & 4 & (0/0) & (0/0)\\ 
7 & 3 & 1 & 4 & (0/0) & (0/0)\\ 
6 & 3 & 2 & 4 & (0/0) & (0/0)\\ 
7 & 3 & 2 & 4 & (0/0) & (0/0)\\ 
\end{tabular}
  \end{center}
\caption{Results for (no own / one opponent) memory, if strategies also play against themselves. ``First it.'' denotes the iteration round, at which the results remain the same until iteration round 1000. TFT wins the tournament, if $b\leq1$ (regardless of $a$), while a comparison of the whole set of strategies is always won by ALLD (defect always).}
\label{tab:1-0}
\end{table}

\begin{table}
\begin{center}
\begin{tabular}[htbp]{ccc|c|c|c}
$T$ &$R$&$P$& First it. &  $G_i^1$ &  Tournament\\ \hline
3 & 2 & 1   & 8    & (0/0) & (0/0) \\
4 & 3 & 1   & 8    & (0/0) & (0/0) \\
5 & 3 & 2   & 12   & (0/0) & (0/0) \\
5 & 4 & 2   & 162 (2) & (0/0) & (0/2), {\it(1/2)} \\
6 & 5 & 2   & 179 (2) & (0/0) & (0/2), {\it(1/2)} \\
4 & 3 & 2   & 108 (2) & (0/0) & (0/0), {\it(0/2)} \\
6 & 4 & 3   & 168 (2) & (0/0) & (0/0), {\it(0/2)} \\
6 & 5 & 3   & 80 (2) & (0/0) & (0/0), {\it(0/2)} \\
5 & 3 & 1   & 4    & (0/0) & (0/0) \\
7 & 4 & 2   & 7    & (0/0) & (0/0) \\
9 & 5 & 3   & 8    & (0/0) & (0/0) \\
4 & 2 & 1   & 4    & (0/0) & (0/0) \\
5 & 2 & 1   & 4    & (0/0) & (0/0) \\
6 & 3 & 1   & 4    & (0/0) & (0/0) \\
7 & 3 & 1   & 4    & (0/0) & (0/0) \\
6 & 3 & 2   & 8    & (0/0) & (0/0) \\
7 & 3 & 2   & 4    & (0/0) & (0/0) \\
\end{tabular}
  \end{center}
\caption{Results for (no own / one opponent) memory, if strategies do not play against themselves. Numbers in brackets in column ``First it.'' denote period length, if results oscillate. Entries marked in italics each second iteration do not co-win the tournament, if the results alternate. This setting is much less prone to lead to co{\"o}peration than if strategies also do play against themselves.}
\label{tab:1-1}
\end{table}

\subsection{Results for One-Own-One-Opponent Memory}
With this configuration beginning with the second iteration step strategies base their decision on two bits, one (the higher bit) in which is encoded the action of their opponent and one in which is remembered their own action. For an overview in Table \ref{tab:11} numbers and behaviors are compared.

For this and all further settings 10,000 iterations (and in special cases more) have been simulated. Results are shown in tables \ref{tab:11-0} and \ref{tab:11-1}.

\begin{table}
\begin{center}
\begin{tabular}[htbp]{c|cc}
numbers for strategies & latest own & latest opponent \\ \hline
(?/1)      & D & D \\
(?/2)      & C & D \\
(?/4)      & D & C \\
(?/8)      & C & C \\   
\end{tabular}
  \end{center}
\caption{A strategy co{\"o}perates, if its number is composed of the elements of this table. TFT for example is (1/12) (co{\"o}perate, if line 3 or line four of this table is remembered: (1/4+8)).}
\label{tab:11}
\end{table}

\begin{table}
\begin{center}
\begin{tabular}[htbp]{ccc|c|c|c}
$T$ &$R$&$P$& First it. &  $G_i^1$ &  Tournament \\ \hline
3 & 2 & 1 & 8 & set of 4 & (1/8), (1/12)\\
4 & 3 & 1 & 66 & (1/8) & (1/8), (1/9), (1/12), (1/13)\\
5 & 3 & 2 & 18 & set of 4 & (1/8), (1/12)\\
5 & 4 & 2 & 21 & (1/8) & (1/8), (1/12)\\
6 & 5 & 2 & 18 & (1/8) & (1/8), (1/9), (1/12), (1/13) \\
4 & 3 & 2 & 12 & set of 4 & (1/8), (1/12)\\
6 & 4 & 3 & 21 & set of 4 & (1/8), (1/12)\\
6 & 5 & 3 & 27 & (1/8) & (1/8), (1/12)\\
5 & 3 & 1 & 8 & set of 4 & (1/8), (1/12)\\
7 & 4 & 2 & 15 & set of 4 & (1/8), (1/12)\\
9 & 5 & 3 & 18 & set of 4 & (1/8), (1/12)\\
4 & 2 & 1 & 1398 & set of 4 & (1/12)\\
5 & 2 & 1 & 10 & set of 4 & (1/8)\\
6 & 3 & 1 & 30 & set of 4 & (1/8), (1/12)\\
7 & 3 & 1 & 6 & set of 4 & (1/8)\\
6 & 3 & 2 & 645 & set of 4 & (1/12), {\it(1/8)}\\
7 & 3 & 2 & 15 & set of 4 & (1/8)\\
\end{tabular}
  \end{center}
\caption{Results for (one own / one opponent) memory, if strategies also play against themselves. ``set of 4'' consists of four strategies: (0/0), (0/2), (0/8), (0/10). All strategies that win the tournament co{\"o}perate in the first iteration and at least continue to co{\"o}perate upon mutual co{\"o}peration (1/ $\geq 8$). (1/12) (TFT) is not among the winners, if $b > a + 1 $. (?/9) is the strategy that sticks with its behavior, if the opponent has co{\"o}perated and else changes it, i.e. it is ``pavlov''. (1/8) can also be seen as a pavlovian strategy, but a more content one than (1/9) -- happy with anything than $S$ and thus repeating the previous behavior except if having received $S$. No rule is among the winners, that continues co{\"o}peration, if the opponent has defected. (Strategy (0/2) would do so, but it never can reach the state that it co{\"o}perates.)}
\label{tab:11-0}
\end{table}

\begin{table}
\begin{center}
\begin{tabular}[htbp]{ccc|c|c|c}
$T$ &$R$&$P$& First it. &  $G_i^1$ &  Tournament \\ \hline
3 & 2 & 1 & 34 & set of 4 & (1/8), (1/12) \\
4 & 3 & 1 & 29 & (1/8) & (1/8), (1/12) \\
5 & 3 & 2 & 30 & set of 4 & (1/8), (1/12) \\
5 & 4 & 2 & 42 & (1/8) & (1/8), (1/12) \\
6 & 5 & 2 & 18 & (1/8) & (1/8), (1/12) \\
4 & 3 & 2 & 23 & set of 4 & (1/8), (1/12) \\
6 & 4 & 3 & 39 & set of 4 & (1/8), (1/12) \\
6 & 5 & 3 & 53 & (1/8) & (1/8), (1/12) \\
5 & 3 & 1 & 363 & set of 4 & (1/8), (1/12) \\
7 & 4 & 2 & 57 & set of 4 & (1/8), (1/12) \\
9 & 5 & 3 & 163 & set of 4 & (1/12), {\it(1/8)} \\
4 & 2 & 1 & 49 & set of 4 & set of 4 \\
5 & 2 & 1 & 69 & set of 4 & set of 4 \\
6 & 3 & 1 & 9 & set of 4 & set of 4 \\
7 & 3 & 1 & 7 & set of 4 & set of 4 \\
6 & 3 & 2 & 66 & set of 4 & set of 4, {\it(0/4)} \\
7 & 3 & 2 & 141 & set of 4 & set of 4 altern. ((0/4), (1/4))\\
\end{tabular}
  \end{center}
\caption{Results for (one own / one opponent) memory, if strategies do not play against themselves. ``set of 4'' consists of four strategies: (0/0), (0/2), (0/8), (0/10).}
\label{tab:11-1}
\end{table}

\subsection{Results for No-Own-Two-Opponent Memory}
Now 10,000 iteration steps were carried out. Again this is far more than the largest number of iterations before the process settles down in some way. Now TFT is (1/2/12) and TF2T is (1/3/14).  Results are shown in tables \ref{tab:2-0} and \ref{tab:2-1}.

\begin{table}
\begin{center}
\begin{tabular}[htbp]{ccc|r|c|c}
$T$ &$R$&$P$& First it. &  $G_i^1$ &  Tournament \\ \hline
3 & 2 & 1   &  383 & (1/2/2)& (1/2/10), (1/3/10),\\
  &   &     &      &        & (1/2/12), (1/3/12)\\
4 & 3 & 1   &  350 & (1/2/2)& (1/3/10), (1/2/14)\\
5 & 3 & 2   &  179 & (0/0/2)& (1/2/10), (1/3/10),\\
  &   &     &      &        & (1/2/12), (1/3/12)\\
5 & 4 & 2   &  422 & (1/2/2)& (1/2/10), (1/3/10), (1/2/12), \\
&&&&&                         (1/3/12), (1/2/14), (1/3/14) \\
6 & 5 & 2   &  397 & (1/2/2)& (1/2/10), (1/3/10), (1/2/12), \\
&&&&&                         (1/3/12), (1/2/14), (1/3/14) \\
4 & 3 & 2   &   53 & (0/0/0)& (1/2/10), (1/3/10), (1/2/12), \\
&&&&&                         (1/3/12), (1/2/14), (1/3/14) \\
6 & 4 & 3   &   35 & (0/0/0)& (1/2/8), (1/3/8), (1/2/10), \\
&&&&&                         (1/3/10), (1/2/12), (1/3/12) \\
6 & 5 & 3   & 1076 & (0/0/0)& (1/3/10) \\
5 & 3 & 1   &  215 & (1/2/2)& (0/3/2)\\
7 & 4 & 2   &  527 & (1/2/2)& (0/3/2)\\
9 & 5 & 3   & 2123 & (1/2/2)& (1/3/10), (1/3/12)\\
4 & 2 & 1   &  719 (2) & (1/2/2)& (1/2/4) altern. (0/3/4)\\
5 & 2 & 1   & 1283 (2) & (0/0/2)& (0/2/4) altern. (0/3/4)\\
6 & 3 & 1   &  299 (4) & (1/2/2)& (1/0/2)\\
7 & 3 & 1   &  395 (4) & (1/2/2)& (1/0/2)\\
6 & 3 & 2   &   41 (2) & (0/0/2)& (1/2/4) altern. (0/3/4)\\
7 & 3 & 2   &  127 (2) & (0/0/2)& (1/2/4) altern. (0/3/4)\\
\end{tabular}
  \end{center}
\caption{Results for (no own / two opponent) memory, if strategies also play against themselves. For 6-3-1 (1/0/2) wins two iteration rounds and then (0/1/2) and then (0/1/2) and (0/3/2) win. For 7-3-1 it is similar, but (0/3/2) does never win. Compared to Table \ref{tab:1-0} TFT (1/2/12) (or even more co{\"o}perative strategies) mostly reappears, only disappears as winner of the tournament for 6-5-3, but newly wins 9-5-3. Thus, the general tendency that payoff matrices with $b\leq1$ produce more co{\"o}peration is kept, but softened. The most co{\"o}perative strategy co-winning a tournament is (1/3/14), which only defects, if it remembers two defections of the opponent. Overall -- compared to the settings with smaller memory -- the dominance of ``always defect'' has vanished, especially in the first round of the tournament.}
\label{tab:2-0}
\end{table}

\begin{table}
\begin{center}
\begin{tabular}[htbp]{ccc|r|c|c}
$T$ &$R$&$P$& First it. &  $G_i^1$ &  Tournament \\ \hline
3 & 2 & 1   &   959 & (1/2/2)& (1/3/10), (1/3/12)\\
4 & 3 & 1   &   219 & (1/2/2)& (1/3/10), (1/3/12), (1/3/14)\\
5 & 3 & 2   &   179 & (0/0/2)& (1/3/10), (1/3/12)\\
5 & 4 & 2   &   720 & (1/2/2)& (1/3/10)\\
6 & 5 & 2   &   619 & (1/2/2)& (0/3/14)\\
4 & 3 & 2   &   276 & (0/0/0)& (1/2/10), (1/3/10), (1/2/12), \\
  &   &     &       &        & (1/3/12), (1/2/14), (1/3/14) \\
6 & 4 & 3   &    38 & (0/0/0)& (1/2/8), (1/3/8), (1/2/10), \\
  &   &     &       &        & (1/3/10), (1/2/12), (1/3/12) \\
6 & 5 & 3   &   422 & (0/0/0)& (1/3/10), (1/0/12), (0/3/14) \\
5 & 3 & 1   &   359 & (1/2/2)& (0/3/2)\\
7 & 4 & 2   & 1224 (3) & (0/0/2)& (1/2/4), {\it(0/2/4)}\\
9 & 5 & 3   & 1644 (3) & (0/0/2)& (1/2/4), {\it(0/2/4)} \\
4 & 2 & 1   & 2891 (2) & (0/0/2)& (0/2/4), {\it((1/2/4) alt. (0/3/4))}\\
5 & 2 & 1   &   13 (2) & (0/0/2)& (0/2/4), {\it(0/3/4)} \\
6 & 3 & 1   &  515 (4) & (1/2/2)& (1/0/2) \\
7 & 3 & 1   &      731 & (1/2/2)& (1/0/2)\\
6 & 3 & 2   &   85 (2) & (0/0/2)& (0/2/4), {\it((1/2/4) alt. (0/3/4))}\\
7 & 3 & 2   &  115 (2) & (0/0/2)& (0/2/4), {\it((1/2/4) alt. (0/3/4))}\\
\end{tabular}
  \end{center}
\caption{Results for (no own / two opponent) memory, if strategies do not play against themselves. For payoff 7-4-2 and 9-5-3 (0/2/4) co-wins in 2 out of 3 rounds. The comparison to Table \ref{tab:1-1} reveals that increasing memory size makes co{\"o}perative strategies much more successful for almost all payoff matrices. None of the payoff matrices that produced oscillating results with size-one memory do so with size-two memory and vice versa.}
\label{tab:2-1}
\end{table}

\subsection{Results for One-Own-Two-Opponent Memory}
In this case, one could in principle reduce the size of the strategy, as it makes no sense to distinguish between strategies that co{\"o}perate or defect in the second iteration, if hypothetically they co{\"o}perate in the first iteration, when in fact they defect in the first iteration. For the simulation the number of strategies has not been reduced to the subset of distinguishable ones, as this would have been a source of error for the source code, and at this stage, the effect on required resources for computation is negligible. Thus for each strategy there are three more that yield exactly the same results against each of the strategies. In the table of results (table \ref{tab:12-0}) just one of the four equivalent strategies is given -- the one with the smallest number. This means that in case of initial defection adding 2, 8, or 10 to the middle number gives the equivalent strategies and in case of initial co{\"o}peration, it is 1, 4, or 5. Therefore TFT is (1/8/240), (1/9/240), (1/12/240), and/or (1/13/240). Even when the results are reduced by naming only one of four strategies linked in this way, this is the first configuration, where the results are too complicated to be understandable at a glance.

There are even more strategies that yield identical results in any combination with any other player: for all strategies that continue to defect (co{\"o}perate) on own defection (co{\"o}peration) those elements of the strategy that determine what to do, following an own co{\"o}peration (defection) are never applied and the value of these elements has no effect. This phenomenon leads to a large number of strategies winning the tournament. Interestingly for some of the payoff matrices the number of winners is smaller around 20 or 30 iterations than at larger numbers of iterations.

For this memory configuration there is almost no difference in the results, if strategies play against themselves or not: the strategies with the most points in the first round of the tournament, and the number of strategies winning the tournament are the same in both cases. Only if the number of strategies winning the tournament is large, a small number of strategies might be exchanged and the iteration round, when the results are stable, is different. In iteration rounds before stability, there can be larger differences, however. We refrain from giving a result table for the case when strategies do not play against themselves.

\begin{table}
\begin{center}
\begin{tabular}[htbp]{ccc|r|c|c}
$T$ &$R$&$P$& First it. &  $G_i^1$  &  Tournament \\ \hline
3 & 2 & 1   &   1436    & set of 4& set of 22, set of 17\\
4 & 3 & 1   &    998    & set of 4& set of 22, set of 20\\
5 & 3 & 2   &    134    & set of 4& set of 22, set of 13\\
5 & 4 & 2   &    234    & set of 4& set of 22, set of 19\\
6 & 5 & 2   &    804    & (1/10/160)& set of 22, set of 39\\
4 & 3 & 2   &   1838    & set of 4& set of 22, set of 37\\
6 & 4 & 3   &    794    & set of 4& set of 22, set of 30  \\
6 & 5 & 3   &    929    & (1/10/160)& set of 22, set of 25 \\
5 & 3 & 1   &   2188    & set of 4& set of 22, (1/10/148)\\
7 & 4 & 2   &     39    & set of 4& set of 22, (1/10/148)\\
9 & 5 & 3   &     45    & set of 4& set of 22, (1/10/148)\\
4 & 2 & 1   &    412    & set of 4& (0/1$\vee$5/180$\vee$244),\\
  &   &     &           &         & (0/5/176$\vee$244)\\
5 & 2 & 1   &    278    & set of 4& (0/1/180) \\
6 & 3 & 1   &    133 (2)& set of 4& (0/1$\vee$5/180$\vee$244),  \\
  &   &     &           &         & (0/5/176$\vee$244), {\it(0/1/244)}\\
7 & 3 & 1   &   2174    & set of 4& (0/1/180)\\
6 & 3 & 2   &    324    & set of 4& (0/1$\vee$5/180$\vee$244),\\
  &   &     &           &         & (0/5/176$\vee$244)\\
7 & 3 & 2   &    422    & set of 4& (0/1/180)\\
\end{tabular}
  \end{center}
\caption{Results for (one own / two opponent) memory, if strategies also play against themselves. The ``$\vee$'' is used as the common meaning of ``or''. (1/10/160) co{\"o}perates in the first and second iteration and then continues to co{\"o}perate, if both strategies have co{\"o}perated, else it defects. This implies that it does not make use of the second to last iteration and is thus simpler than possible. Except for the definite co{\"o}peration in the second iteration, it is strategy (1/8) from the (one / one) setting. ``set of 4'' consists of (0/0/1$\vee$9$\vee$129$\vee$137), which all do make use of the information on the opponent's second to last action. ``set of 22'' is (1/8$\vee$10/176$\vee$180$\vee$208$\vee$212$\vee$240$\vee$244), (1/8/144$\vee$146$\vee$148$\vee$150$\vee$178$\vee$182$\vee$210$\vee$214$\vee$242$\vee$246) and by this includes TFT. ``set of 13'' is (1/10/148), (1/8$\vee$10/132$\vee$140$\vee$164$\vee$196$\vee$204$\vee$228). ``set of 17'' includes ``set of 13'', (1/8/168$\vee$172$\vee$232), and (1/10/144). ``set of 30'' contains ``set of 13'', (1/8$\vee$/10/128$\vee$136$\vee$160$\vee$192$\vee$200$\vee$224), (1/8/130$\vee$162$\vee$194$\vee$226), and (1/10/144). ``set of 37'' consists of ``set of 30'', (1/8$\vee$10/168$\vee$172$\vee$232), and (1/10/236). The remaining four sets (``set of 20'', ``set of 39'', ``set of 25'', and ``set of 29'') share in common (1/10/168$\vee$172$\vee$184$\vee$188$\vee$204$\vee$232$\vee$236$\vee$248$\vee$252), which includes TF2T. A total of 41 further strategies appear as members of these sets, of which a majority (28) have not appeared earlier in this table and table's caption.
}
\label{tab:12-0}
\end{table}

\subsection{Results for Two-Own-One-Opponent Memory}
This configuration is interesting as one can interpret a strategy considering a remembered opponent's action as reaction to an as well remembered own action. While TFT is (1/8/240), a strategy additionally co{\"o}perating in such a case would be (1/8/244). As Table \ref{tab:21-0} shows, sometimes only TFT appears among the winners of the tournament, sometimes both these strategies. Only with payoff matrix 6-5-2 the more forgiving strategy wins but not TFT. It is the more tricky strategy (1/8/228) that applies this kind of forgiveness, which is more successful than TFT.

In this setting as well, it has only minor effects if a strategy plays against itself or not. Therefore the results for the case when they do not is omitted.

\begin{table}
\begin{center}
\small
\begin{tabular}[htbp]{ccc|r|c|c}
$T$ &$R$&$P$& First it. &  $G_i^1$  &  Tournament \\ \hline
3 & 2 & 1   & 539 (2) & (0/1/4)& (1/8$\vee$10/164$\vee$228), \\
  &   &     &         &        & (1/10/set of 13 altern. set of 14)\\
4 & 3 & 1   & 338     & (0/1/4)& (1/8/228$\vee$229)\\
5 & 3 & 2   & 8367    & (0/1/4)& (1/8/228)\\
5 & 4 & 2   &  107    & (0/1/4)& (1/8$\vee$10/224$\vee$228$\vee$240$\vee$244),\\
  &   &     &         &        & (1/10/set of 14)\\
6 & 5 & 2   &  111    & (0/1/4)& (1/8/228$\vee$229$\vee$244)\\
4 & 3 & 2   & 3768    & (0/1/4)& (1/8$\vee$10/224$\vee$228$\vee$240),\\
  &   &     &         &        & (1/10/set of 11)\\
6 & 4 & 3   & 242     & (0/1/4)& (1/8$\vee$10/164$\vee$224$\vee$228$\vee$240),\\
  &   &     &         &        & (1/10/set of 12)\\
6 & 5 & 3   & 483     & (0/1/4)& (1/8/224$\vee$228$\vee$240$\vee$244)\\
5 & 3 & 1   & 106     & (0/1/4)& (1/8$\vee$10/164$\vee$228), \\
  &   &     &         &        & (1/10/160$\vee$161$\vee$176$\vee$177$\vee$224$\vee$225$\vee$240$\vee$241)\\
7 & 4 & 2   & 5989 (4)& (0/1/4)& (1/{\it 8}$\vee$10/160$\vee$176$\vee$224), (1/10/240)\\
9 & 5 & 3   & 350     & (0/1/4)& (1/8$\vee$10/160$\vee$176$\vee$224), (1/10/240)\\
4 & 2 & 1   & 32 (2)  & (0/1/4)& (1/8/224$\vee${\it 160}$\vee${\it 176})\\
5 & 2 & 1   & 32      & (0/1/4)& (0/5/224)\\
6 & 3 & 1   & 407 (2) & (0/1/4)& (1/8/160$\vee$161$\vee$176$\vee$177$\vee$224$\vee$225), \\
  &   &     &         &        & altern. (0/5/224$\vee$225)\\
7 & 3 & 1   &   29    & (0/1/4)& (0/5/224$\vee$225)\\
6 & 3 & 2   & 37 (2)  & (0/1/4)& (1/8/160$\vee$176$\vee$224),} altern. {\it (0/5/224)\\
7 & 3 & 2   &   35    & (0/1/4)& (0/5/224)\\
\end{tabular}
\normalsize
\end{center}
\caption{Results for (two own / one opponent) memory, if strategies also play against themselves. For the payoff matrices from the top down to 5-3-1 strategy (1/8/228) always is among the winners of the tournament. It is the strategy that almost plays tit for tat, but does not co{\"o}perate, if the opponent has co{\"o}perated and itself has defected two times, but does co{\"o}perate, if the opponent has defected after itself has defected, even, if itself has co{\"o}perated in the most recent game. For the winner strategy (0/1/4) of the first round of the tournament, this history is even the only case when it co{\"o}perates. For the payoff matrices 5-3-2 and 7-4-2 20,000 iterations were calculated to verify the late stability, respectively period 4.}
\label{tab:21-0}
\end{table}

\subsection{Results for No-Own-Three-Opponent Memory}
Regarding the number of strategies this setting is the largest investigated in this work. The number of iterations until the results settle varies greatly among the various payoff matrices. In fact for some payoff matrices they did not stabilize before the 30,000th iteration. At this point we refrained from performing further calculations and accepted the (non-)result as open issue for future investigations. However, even for payoff matrices with which stable results appear to have been reached it cannot be excluded that after some 10,000 iterations more different winners would result, as in the more volatile cases. Another surprising observation was that the results sometimes appeared to have reached a final state but then started changing again. After all, for remembering one opponent's action, stable results appeared after approximately 10 iterations, for remembering two opponents' moves it was about 1,000 iterations. So, it is not unrealistic to assume that remembering three opponents' actions may need 100,000 or even more iterations until the results do not change anymore. 

Further difficulties may arise from precision issues in the calculation. During the tournament it is decided by comparison with the average of points, if the strategies may participate in the next round. The average is calculated by dividing one very large number by another very large number. As a consequence the size comparison between average and individual result may be faulty, if in fact a strategy has exactly achieved the average of points and by this kicked out of the tournament. Another resource problem is the possibility that the sum of points produces an overflow in the corresponding integer variable. That such considerations could be relevant when dealing with such large numbers is based on general experience with complex simulations; in the results there was no explicit hint that such issues really occured, except for that the long instability of results that appeared to be surprising in principle could be attributed to them. Ruling them out would need a second computer system with different hardware architecture or a very thorough understanding of the CPU and the compiler that were used. None of these were sufficiently available. Additionally one has to consider that each simulation run currently takes days to arrive at the number of iterations where these issues could be relevant. In a nutshell: using up-to-date standard computer systems the no-own-three-opponent-memory case today is at the edge of what is accessible. Definately ruling out negative effects that falsify the results and doing this with maintainable effort remains to be done in the future.

As calculating the payoff and evaluating the tournament takes more computation time than calculating the results of the dilemma, beyond 10,000 iterations only for the last one hundred iterations ahead of the full thousands payoff and tournament were calculated. This in turn implies that the iteration number after which the results did not change anymore can only be given approximately. 

Having said all this, it becomes obvious that the results of this section need to be considered as preliminary -- the more the later the assumed stability was observed.

A different problem is that in some cases the number of winners of the tournament is too large to give all of the winning strategies in this paper. However, the remaining cases should be sufficient to demonstrate the type and especially variants of strategies winning the tournament

A majority of the strategies winning the first round of the tournament co{\"o}perate, when the earliest opponent's action they remember was co{\"o}peration and any other defection. This is a trend which was already present with the one element smaller memory, but it was not as pronounced. This strategy is interesting in a sense as it uses the last chance to avoid breaking entirely with the opponent. To find a catchy name for this strategy, recall Mephisto's behavior toward God in the Prologue in Heaven of Faust I: {\em ``The ancient one I like sometimes to see, And not to break with him am always civil''}\footnote{The German original ``Von Zeit zu Zeit seh ich den Alten gern, und h{\"u}te mich mit ihm zu brechen.'' even more stresses the occasional character of the co{\"o}perative interaction.}, where even considering all the competition between the two, Mephisto avoids entirely abandoning co{\"o}peration. If one extrapolates {\em Mephisto} to even larger memory sizes, co{\"o}peration vanishes more and more, although there is some basic co{\"o}perative tendency kept in the strategy. There are two questions: if this trend would actually continue infinitely, when memory size is increased further, and what it means that for example the case of one-own-two-opponent-memory-size yields strategies as winners of the first round of the tournament that have entirely different characteristics.

The results are shown in table \ref{tab:3-0}.

\begin{table}
\begin{center}
\small
\begin{tabular}[htbp]{ccc|r|c|c}
$T$ &$R$&$P$& First it.  			 &  $G_i^1$  &  Tournament \\ \hline
3 & 2 & 1   & $\approx$ 24,000 & (1/2/2/2) & 138 strategies\\
4 & 3 & 1   & $\approx$ 27,000 & (0/0/0/9) & (1/0/10/246), (1/0/14/230),\\
  &   &     &                  &           & (1/0/11/230$\vee$246), (1/0$\vee$1/14/236$\vee$246), \\
  &   &     &                  &           & (1/0$\vee$1/15/228$\vee$230$\vee$236$\vee$246)\\
5 & 3 & 2   & $\approx$ 9,000  & (0/0/0/2) & 117 strategies e.g.\\ 
  &   &     &                  &           & (1/2$\vee$3/12$\vee$13$\vee$14$\vee$15/162$\vee$164$\vee$228$\vee$240)\\
5 & 4 & 2   &       -          & (1/2/2/2) & (0/2/7/230), \\
  &   &     &                  &           & (0/0/15/230), (0/2/230$\vee$238)\\
6 & 5 & 2   & $\approx$ 21,000 & (1/2/2/2) & (0/1$\vee$3/10$\vee$11/230$\vee$246$\vee$254),\\
  &   &     &                  &           & (0/0$\vee$1$\vee$2$\vee$3$\vee$8$\vee$9$\vee$10$\vee$11/230$\vee$246$\vee$254)\\
4 & 3 & 2   & $\approx$ 22,000 & (0/0/0/2) & 136 strategies e.g.\\
  &   &     &                  &           & (1/2/12/166), (1/3/8/240), (1/3/15/252)\\
6 & 4 & 3   & $\approx$ 22,000 & (0/0/0/2) & 207 strategies e.g.\\
  &   &     &                  &           & (1/2/12/160), (1/3/8/240), (1/3/15/248)\\
6 & 5 & 3   & $\approx$ 9,000  & (0/0/0/2) & (0/1$\vee$3/10$\vee$11/230$\vee$246$\vee$254),\\
  &   &     &                  &           & (0/3/0$\vee$1$\vee$8$\vee$9$\vee$10$\vee$11/230)\\
5 & 3 & 1   &    678           & (0/0/0/9) & 74 strategies e.g.\\
  &   &     &                  &           & (1/2$\vee$3/12$\vee$13$\vee$14$\vee$15/162$\vee$176$\vee$228$\vee$240)\\
7 & 4 & 2   & $\approx$ 9,000  & (1/2/10/2)& 78 strategies e.g.\\
  &   &     &                  &           & (1/2$\vee$3/12$\vee$13$\vee$14$\vee$15/162$\vee$176$\vee$228$\vee$240)\\
9 & 5 & 3   & $\approx$ 12,000 & (0/0/0/2) & 80 strategies e.g.\\
  &   &     &                  &           & (1/2$\vee$3/12$\vee$13$\vee$14$\vee$15/162$\vee$176$\vee$228$\vee$240)\\
4 & 2 & 1   & 609 (2)          & (0/0/0/2) & (1/3/8$\vee$9/226) alt. (0/3/13$\vee$15/226)\\
5 & 2 & 1   &  1695            & (0/0/0/2) & (0/3/13/226)\\
6 & 3 & 1   &  1923 (2)        & (1/2/10/2)& (1/3/8$\vee$9/226) alt. (0/3/13$\vee$15/226)\\
7 & 3 & 1   & $\approx$ 9,000  & (1/2/10/2)& (0/3/13/226)\\
6 & 3 & 2   & $\approx$ 9,000 (2) & (0/0/0/2) & (0/3/15/226) alt.\\
  &   &     &                  &           & (1/3/8$\vee$9/226), (1/3/9/240)\\
7 & 3 & 2   & 1229             & (0/0/0/2) & (0/3/13/226)\\
\end{tabular}
\normalsize
  \end{center}
\caption{Results for remembering three preceding opponents' actions. (Strategies do play against themselves.) For 5-4-2-0 after a varying number of iterations (roughly 10) another result with 14 tournament winning strategies appears. These do not include the 6 given here}
\label{tab:3-0}
\end{table}

\section{Summary and Outlook}
The calculations of this work reveal a strong dependence of the results of the tournament on the details of the payoff matrix. It is not sufficient to distinguish, if $T + S = R + P$ and $2R > T + S$ hold or not. This means that one has to be careful drawing conclusions, if the prisoner's dilemma is used as a toy model for some real system. Of course, as this work restricted strategies to limited memory size, there might be strategies relying on infinite memory that outperform all of these regardless of the details of the payoff matrix. So, the main result of this work is not that everything changes with a different payoff matrix, but that one should not be too faithful that the precise choice of the payoff matrix is irrelevant.

As expected the two basic relations $T + S = R + P$ and $2R > T + S$ clearly have an influence on the results, as subsets of strategies appear among the winners in tendency depending if these relations hold or if they do not so. The picture is a bit different for the winner of the first round of the tournament, when all strategies still participate: there are fewer strategies appearing as winners, but if there is more than one for a memory configuration, there is no obvious pattern based on these relations that tells which strategy wins if a specific payoff matrix is applied. In total, one cannot claim that the details of the payoff matrix will dominate each element of the results in any case. However, in general one can say that the results do depend on the specific choice of the payoff matrix. Furthermore it is not only not possible to find one generally best or a set of generally best strategies, but -- if one compares the winners of the first round of the tournament and the tournament as a whole -- even for a specific payoff matrix it cannot be decided in general, if co{\"o}perating is a good or bad idea, as this depends on the kind of result that decides about the winner. 

While for these reasons, it is usually not possible to use the prisoner's dilemma as some kind of proof that in some real system co{\"o}perating yields best payoff, the results of this work -- as of a lot of preceding works -- helps to bear in remembrance that co{\"o}perating {\em might} be the better idea, even if at first sight one might have the opposite impression. The iterated prisoner's dilemma obviously is an abstract and simplified model for any real social system and the four entries of the payoff matrix often are not set quantitatively by the real system. In such cases conclusions drawn from calculations can only be valid if the results do not significantly depend on details of the payoff matrix.

In some cases the results stabilized only after a very large number of iterations, a number far larger than for example the number of iterations in the tournaments performed by Axelrod \cite{_Axelrod1985}. This does not necessarily mean that it is useless to investigate cases with fewer iterations, as also before the results stabilize, the results oscillate between two sets or between a set and a proper subset. As the number of iterations for stability grows with the number of participating strategies and as the number of participating strategies is already quite large in cases, when stability only occurs beyond 1000 iterations, one can assume that for most investigations of the iterated prisoner's dilemma that have been published so far, the number of iterations was sufficiently high. Still, the results of this work indicate that an investigation of the effect of having $\pm 20$ iterations usually should be worth the effort.

The results show a tendency that for increased memory size somewhat co{\"o}perative strategies score better. There have been investigations on the dependency of good memory and scoring in an iterated prisoner's dilemma \cite{Milinski1998,Winkler2008}, however, the facing work is rather indifferent on this issue. With memory size also the number of strategies increases and co{\"o}perative strategies find more strategies that co{\"o}perate as well. A comparison of Tables \ref{tab:1-0} and \ref{tab:1-1} supports this idea, as it shows how it benefits co{\"o}perative strategies, when there is one more co{\"o}perative counterpart (themselves) participating in the tournament. The fact that with increasing memory size in the end it does not play any further role, if strategies play themselves or not, shows that in these cases the strategies are related to some of the others, in a way that in effect playing against them is as playing against themselves. On the other hand, if a good memory would not matter then there should be more strategies among the winners that do not make use of principally available more past information.

In this work the results have mainly been presented and -- despite the considerable extent of the paper -- only scarcely been analyzed and discussed. There are plenty of possibilities to discuss the success or poor performance of a specific strategy in a specific memory configuration with a specific payoff matrix in analytical terms. For settings that yield large sets of tournament winners, the results can be investigated statistically. Once stronger computational resources are available larger memories can be investigated and the case of remembering three opponent's actions can be investigated more reliable.

In this work the idea was to simulate as many rounds as are necessary to yield stable results. The development of the results over the rounds was not and thus could be investigated in further studies.

For the tournament itself one can think of many variants. One could for example only eliminate those strategies scoring worst in an iteration, or eliminate always (as far as possible) exactly half of the strategies still running. It is also possible to allow initial population weights different than one.

And finally the role of the payoff matrix can be investigated in greater depth. In this work no two payoff matrices always gave the same result (although the results of 7-4-2 and 9-5-3 were always at least similar). Is it possible at all that two payoff matrices that are not related trivially yield the same results? And if this is the case, what is (if it exists) the simplest parametrization and set of relations between the parameters, which generates all payoff matrices that yield all possible results? Can the winning strategies or the number of iterations until stability be derived analytically? 

The differences between the results with different payoff matrices might as well reduce, if the tournament were not carried out in a binary way, but if the frequency of a strategy could take a real value and frequencies of a round were dependent on the score (fitness) of the preceding round. It would then be possible for a strategy to score below average for example in the first round, but recover in subsequent rounds.

\section{Acknowledgments}
The author is grateful to his company {\em PTV -- Planung Transport Verkehr} for providing the computing hardware and computation time.

\nocite{_Canter1980,_Smelser2001}
\bibliographystyle{utphys5}
\bibliography{CA_PD}

\providecommand{\href}[2]{#2}\begingroup\raggedright\begin{thebibliography}{10}

\bibitem{Flood1958}
M.~Flood, ``{Some experimental games}'',
  \href{http://dx.doi.org/10.1287/mnsc.5.1.5}{{\em Management Science}
  {\bfseries 5} no.~1, (1958) 5--26}.

\bibitem{_Axelrod1985}
R.~Axelrod, {\em {The Evolution of Cooperation}}.
\newblock Basic Books, New~York, NY, 1985.
\newblock ISBN:0-465-02122-0.

\bibitem{Axelrod1981}
R.~Axelrod and W.~Hamilton, ``{The evolution of cooperation}'',
  \href{http://dx.doi.org/10.1126/science.7466396}{{\em Science} {\bfseries
  211} no.~4489, (1981) 1390--1396}.

\bibitem{Sime1980}
J.~Sime, ``{The Concept of Panic}'', in Canter \cite{_Canter1980}, ch.~5,
  pp.~63--81.
\newblock ISBN:978-1853461392.

\bibitem{Keating1982}
J.~Keating, ``{The myth of panic}'', {\em Fire Journal} (1982) 57--62.

\bibitem{Laur1997}
U.~Laur, H.~Jaakula, J.~Metsaveer, K.~Lehtola, H.~Livonen, T.~Karppinen, A.-L.
  Eksborg, H.~Rosengren, and O.~Noord, ``{Final Report on the Capsizing on 28
  September 1994 in the Baltic Sea of the Ro-Ro Passenger Vessel MV Estonia}'',
  tech. rep., The Joint Accident Investigation Commission of Estonia, Finland
  and Sweden, December, 1997.
\newblock \url{http://www.onnettomuustutkinta.fi/estonia/}.

\bibitem{Quarantelli2001}
E.~Quarantelli, ``{The sociology of panic}'', in Smelser and Baltes
  \cite{_Smelser2001}, pp.~11020--11030.
\newblock ISBN:0-080-43076-7.

\bibitem{Clarke2002}
L.~Clarke, ``{Panic: Myth or Reality?}'', {\em contexts} {\bfseries 1} no.~3,
  (2002) . \url{http://www.leeclarke.com/docs/Clarke_panic.pdf}.

\bibitem{Mawson2005}
A.~Mawson, ``{Understanding Mass Panic and Other Collective Responses to Threat
  and Disaster}'', \href{http://dx.doi.org/10.1521/psyc.2005.68.2.95}{{\em
  Psychatry} {\bfseries 68} (2005) 95--113}.

\bibitem{Fahy2005}
R.~Fahy and G.~Proulx, ``{Analysis of Published Accounts of the World Trade
  Center Evacuation}'', tech. rep., National Institute of Standards and
  Technology, 9, 2005.
\newblock \url{http://www.nist.gov/customcf/get_pdf.cfm?pub_id=101422}.

\bibitem{Drury2009}
J.~Drury, C.~Cocking, and S.~Reicher, ``The Nature of Collective Resilience:
  Survivor Reactions to the 2005 London Bombings'', {\em International Journal
  of Mass Emergencies and Disasters} {\bfseries 27} (2009) .
  \url{http://www.sussex.ac.uk/affiliates/panic/IJMED\%20Drury\%20et\%20al.\%2%
02009.pdf}.

\bibitem{McFadden1991}
R.~McFadden, ``{ Stampede at City College; Inquiries Begin Over City College
  Deaths}'', {\em The New York Times} (1991) .
  \url{http://query.nytimes.com/gst/fullpage.html?res=9D0CEED91738F933A05751C1%
A967958260}. 12/31/1991.

\bibitem{Schelajew2000}
J.~Schelajew, E.~Schelajewa, and N.~Semjonow, {\em {Nikolaus II. Der letzte
  russische Zar}}.
\newblock Bechterm\"unz, Augsburg, 2000.
\newblock ISBN:3-82890-270-7.

\bibitem{Kretz2007a}
T.~Kretz,
  \href{http://nbn-resolving.de/urn:nbn:de:hbz:464-20070302-120944-7}{{\em
  {Pedestrian Traffic -- Simulation and Experiments}}}.
\newblock PhD thesis, Universit\"at Duisburg-Essen, 2007.

\bibitem{Schreckenberg2008}
C.~Rogsch, M.~Schreckenberg, E.~Tribble, W.~Klingsch, and T.~Kretz,
  \href{http://dx.doi.org/10.1007/978-3-642-04504-2_72}{``{Was it Panic? An
  Overview about Mass-Emergencies and their Origins all over the World for
  Recent Years}'',} in {\em {Pedestrian and Evacuation Dynamics 2008}},
  W.~Klingsch, C.~Rogsch, A.~Schadschneider, and M.~Schreckenberg, eds.,
  pp.~743--755.
\newblock Springer-Verlag, Berlin Heidelberg, 2010.
\newblock ISBN: 978-3-642-04503-5.

\bibitem{Schadschneider2009}
A.~Schadschneider, W.~Klingsch, H.~Kl{\"u}pfel, T.~Kretz, C.~Rogsch, and
  A.~Seyfried,
  \href{http://dx.doi.org/10.1007/978-0-387-30440-3_187}{``{Evacuation
  Dynamics: Empirical Results, Modeling and Applications}'',} in {\em
  {Encyclopedia of Complexity and Systems Science}}, R.~Meyers, ed.
\newblock Springer, Berlin Heidelberg New York, 2009.
\newblock \href{http://arxiv.org/abs/0802.1620}{{\ttfamily arXiv:0802.1620
  [physics.soc-ph]}}.
\newblock ISBN:978-0-387-75888-6.

\bibitem{_Rapoport1965}
A.~Rapoport and A.~Chammah, {\em {Prisoner's dilemma}}.
\newblock University of Michigan Press, 1965.
\newblock ISBN:978-0472061655.

\bibitem{Trivers1971}
R.~Trivers, ``{The evolution of reciprocal altruism}'',
  \href{http://dx.doi.org/10.1086/406755}{{\em The Quarterly Review of Biology}
  {\bfseries 46} no.~1, (1971) 35}.

\bibitem{_Axelrod1997}
R.~Axelrod, {\em {The Complexity of Cooperation}}.
\newblock Princeton University Press, Princeton, New Jersey, 1997.
\newblock ISBN:0-691-01568-6.

\bibitem{Nowak1993}
M.~Nowak and K.~Sigmund, ``{A strategy of win-stay, lose-shift that outperforms
  tit-for-tat in the Prisoner's Dilemma game}'',
  \href{http://dx.doi.org/10.1038/364056a0}{{\em Nature} {\bfseries 364}
  no.~6432, (1993) 56--58}.

\bibitem{Baek1989}
S.~Baek and B.~Kim, ``{Intelligent tit-for-tat in the iterated prisoner�s
  dilemma game}'', \href{http://dx.doi.org/10.1103/PhysRevE.78.011125}{{\em
  Theory Decis Phys Rev E} {\bfseries 78} (1989) 011125},
  \href{http://arxiv.org/abs/0807.2105}{{\ttfamily arXiv:0807.2105
  [q-bio.PE]}}.

\bibitem{Nowak1992}
M.~Nowak and R.~May, ``{Evolutionary games and spatial chaos}'',
  \href{http://dx.doi.org/10.1038/359826a0}{{\em Nature} {\bfseries 359}
  no.~6398, (1992) 826--829}.

\bibitem{Nowak1993b}
M.~Nowak and R.~May, ``{The spatial dilemmas of evolution}'',
  \href{http://dx.doi.org/10.1142/S0218127493000040}{{\em International Journal
  of Bifurcation and Chaos} {\bfseries 3} (1993) 35--78}.

\bibitem{Grim1997}
P.~Grim, ``{The undecidability of the spatialized prisoner�s dilemma}'',
  \href{http://dx.doi.org/10.1023/A:1004959623042}{{\em Theory and Decision}
  {\bfseries 42} (1997) 53--80}.

\bibitem{Nakamaru1997}
M.~Nakamaru, H.~Matsuda, and Y.~Iwasa, ``{The evolution of cooperation in a
  lattice-structured population}'',
  \href{http://dx.doi.org/10.1006/jtbi.1996.0243}{{\em Journal of Theoretical
  Biology} {\bfseries 184} no.~1, (1997) 65--81}.

\bibitem{Kirchkamp2000}
O.~Kirchkamp, ``{Spatial evolution of automata in the prisoners� dilemma}'',
  \href{http://dx.doi.org/10.1016/S0167-2681(00)00118-9}{{\em Journal of
  Economic Behavior and Organization} {\bfseries 43} no.~2, (2000) 239--262}.

\bibitem{Schweitzer2002}
F.~Schweitzer, L.~Behera, and H.~M{\"u}hlenbein, ``{Evolution of Cooperation in
  a Spatial Prisoner's Dilemma}'',
  \href{http://dx.doi.org/10.1142/S0219525902000584}{{\em Advances in Complex
  Systems} {\bfseries 5} no.~2-3, (2002) 269--299},
  \href{http://arxiv.org/abs/cond-mat/0211605}{{\ttfamily
  arXiv:cond-mat/0211605 [cond-mat.stat-mech]}}.

\bibitem{Masuda2003}
N.~Masuda and K.~Aihara, ``{Spatial prisoner's dilemma optimally played in
  small-world networks}'',
  \href{http://dx.doi.org/10.1016/S0375-9601(03)00693-5}{{\em Physics Letters
  A} {\bfseries 313} no.~1-2, (2003) 55--61}.

\bibitem{Fort2005}
H.~Fort and S.~Viola, ``{Spatial patterns and scale freedom in Prisoner's
  Dilemma cellular automata with Pavlovian strategies}'',
  \href{http://dx.doi.org/10.1088/1742-5468/2005/01/P01010}{{\em J. Stat. Mech}
  {\bfseries 01010} (2005) }, \href{http://arxiv.org/abs/0412737}{{\ttfamily
  arXiv:0412737 [cond-mat.stat-mech]}}.

\bibitem{Alonso2006}
J.~Alonso, A.~Fern{\'a}ndez, and H.~Fort, ``{Prisoner's Dilemma cellular
  automata revisited: evolution of cooperation under environmental pressure}'',
  \href{http://dx.doi.org/10.1088/1742-5468/2006/06/P06013}{{\em J. Stat. Mech}
  (2006) P06013}, \href{http://arxiv.org/abs/0512187}{{\ttfamily arXiv:0512187
  [physics.soc-ph]}}.

\bibitem{Holland1992}
J.~Holland, ``{Genetic Algorithms Computer programs that ``evolve'' in ways
  that resemble natural selection can solve complex problems even their
  creators do not fully understand}'', {\em Scientific American} {\bfseries
  267} (1992) 66--72.

\bibitem{Michalewicz1996}
Z.~Michalewicz, {\em {Genetic algorithms + data structures = evolution
  programs}}.
\newblock Springer, 3~ed., 1996.
\newblock ISBN:978-3540606765.

\bibitem{Salhi1996}
A.~Salhi, H.~Glaser, D.~Roure, and J.~Putney, ``{The Prisoners' Dilemma
  Revisited}''. Eprint, 1996.

\bibitem{Doebeli2005}
M.~Doebeli and C.~Hauert, ``{Models of cooperation based on the Prisoner's
  Dilemma and the Snowdrift game}'',
  \href{http://dx.doi.org/10.1111/j.1461-0248.2005.00773.x}{{\em Ecology
  Letters} {\bfseries 8} no.~7, (2005) 748--766}.

\bibitem{Kuhn2007}
S.~Kuhn, ``{Prisoner's dilemma}'', in {\em Stanford Encyclopedia of
  Philosophy}, E.~Zalta, ed.
\newblock The Metaphysics Research Lab, Center for the Study of Language and
  Information, Stanford University, Stanford, CA, 2007.
\newblock \url{http://plato.stanford.edu/entries/prisoner-dilemma/}.

\bibitem{Moore1956}
E.~Moore, ``{Gedanken-experiments on sequential machines}'', {\em Automata
  studies} {\bfseries 34} (1956) 129--153.

\bibitem{Abreu1988}
D.~Abreu and A.~Rubinstein, ``{The structure of Nash equilibrium in repeated
  games with finite automata}'', {\em Econometrica: Journal of the Econometric
  Society} {\bfseries 56} no.~6, (1988) 1259--1281.
  \url{http://arielrubinstein.tau.ac.il/papers/29.pdf}.

\bibitem{Linster1992}
B.~Linster, ``{Evolutionary Stability in the Infinitely Repeated Prisoners'
  Dilemma Played by Two-State Moore Machines}'', {\em Southern economic
  journal} {\bfseries 58} no.~4, (1992) .

\bibitem{Miller1996}
J.~Miller, ``{The coevolution of automata in the repeated prisoner's
  dilemma}'', \href{http://dx.doi.org/10.1016/0167-2681(95)00052-6}{{\em
  Journal of Economic Behavior and Organization} {\bfseries 29} no.~1, (1996)
  87--112}.

\bibitem{_Wolfram1986}
S.~Wolfram, {\em {Theory and Application of Cellular Automata}}.
\newblock World Scientific, Singapore, 1986.
\newblock ISBN:9-971-50123-6.

\bibitem{_Wolfram2002}
S.~Wolfram, {\em {A New Kind of Science}}.
\newblock Champaign, IL: Wolfram Media, 2002, 2002.
\newblock \url{http://www.wolframscience.com/nksonline/toc.html}.
\newblock ISBN:1-579-55008-8.

\bibitem{Beaufils1998}
B.~Beaufils, J.~Delahaye, and P.~Mathieu,
  \href{http://dx.doi.org/10.1007/BFb0040757}{``{Complete classes of strategies
  for the classical iterated prisoner's dilemma}'',} in {\em Evolutionary
  Programming VII: 7th International Conference, EP'98, San Diego, California,
  USA, March 25-27, 1998: Proceedings}, W.~D. Porto~V.W., Saravanan~N., ed.,
  pp.~33--42.
\newblock Springer, 1998.
\newblock ISBN:3-540-64891-7.

\bibitem{Hauert1997}
C.~Hauert, ``{Effects of increasing the number of players and memory size in
  the iterated Prisoner's Dilemma: a numerical approach}'',
  \href{http://dx.doi.org/10.1098/rspb.1997.0073}{{\em Proceedings of the Royal
  Society B: Biological Sciences} {\bfseries 264} no.~1381, (1997) 513--519}.

\bibitem{Townsley2006}
M.~Townsley, M.~Weeks, R.~Ragade, and A.~Kumar, ``{A Large Scale, Distributed,
  Iterated Prisoner's Dilemma Simulation}'', {\em Transactions on Advanced
  Research} {\bfseries 2} no.~2, (2006) 58--63.
  \url{http://internetjournals.net/journals/tar/2006/July/Paper\%2010.pdf}.

\bibitem{Price1970}
G.~Price, ``{Selection and Covariance}'',
  \href{http://dx.doi.org/10.1038/227520a0}{{\em Nature} {\bfseries 227} (1970)
  520--521}.

\bibitem{Frank1995}
S.~Frank, ``{George Price's Contributions to Evolutionary Genetics}'',
  \href{http://dx.doi.org/10.1006/jtbi.1995.0148}{{\em J. theor. Biol.}
  {\bfseries 175} (1995) 373--388}.

\bibitem{Milinski1998}
M.~Milinski and C.~Wedekind, ``{Working memory constrains human cooperation in
  the Prisoner's Dilemma}'',
  \href{http://dx.doi.org/10.1073/pnas.95.23.13755}{{\em PNAS} {\bfseries 95}
  no.~23, (1998) 13755--13758}.

\bibitem{Winkler2008}
I.~Winkler, K.~Jonas, and U.~Rudolph, ``{On the Usefulness of Memory Skills in
  Social Interactions}'',
  \href{http://dx.doi.org/10.1177/0022002707312606}{{\em Journal of Conflict
  Resolution} {\bfseries 52} no.~3, (2008) 375--384}.

\bibitem{_Canter1980}
D.~Canter, ed., {\em Fires and Human Behaviour}, vol.~1.
\newblock John Wiley \& Sons Ltd., London, 1980.
\newblock ISBN:978-1853461392.

\bibitem{_Smelser2001}
N.~Smelser and P.~Baltes, eds., {\em {International Encyclopedia of the Social
  and Behavioral Sciences}}.
\newblock Elsevier, Oxford, 2001.
\newblock ISBN:0-080-43076-7.

\end{thebibliography}\endgroup
\end{document}